\documentclass[sigconf,screen, authorversion]{acmart}
\PassOptionsToPackage{table,xcdraw}{xcolor}
\AtBeginDocument{%
  }

\copyrightyear{2025}
\acmYear{2025}
\setcopyright{rightsretained}
\acmConference[CHI EA '25]{Extended Abstracts of the CHI Conference on Human Factors in Computing Systems}{April 26-May 1, 2025}{Yokohama, Japan}
\acmBooktitle{Extended Abstracts of the CHI Conference on Human Factors in Computing Systems (CHI EA '25), April 26-May 1, 2025, Yokohama, Japan}\acmDOI{10.1145/3706599.3720080}
\acmISBN{979-8-4007-1395-8/2025/04}

\acmSubmissionID{2287}

\graphicspath{{figures/}{pictures/}{images/}{./}} 

\usepackage{amsmath}
\usepackage{algorithm}

\usepackage{color}
\usepackage{bbding}
\usepackage{booktabs}

\usepackage{enumitem}
\setlist{noitemsep,parsep=0pt,partopsep=0pt, leftmargin=10pt} 
\usepackage{listings}
\usepackage{multirow}
\usepackage{colortbl}

\definecolor{wyf}{RGB}{237, 85, 106} 

\newcommand{\eg}{{\it e.g.,\ }}
\newcommand{\etal}{{\it et al.\ }}
\newcommand{\etc}{{\it etc.}}
\newcommand{\ie}{{\it i.e.,\ }}
\newcommand{\bpstart}[1]{{\textbf{#1.}}}
\newcommand{\nistart}[1]{{\noindent\textbf{#1.}}}
\newcommand{\transparentcolorbox}[2]{\colorbox{#1!90}{#2}}

\definecolor{Restrict}{HTML}{eee0da}
\definecolor{Expand}{HTML}{d3e5ef}
\definecolor{Refine}{HTML}{dbeddb}
\definecolor{Organize}{HTML}{e8deee}
\definecolor{Follow-up}{HTML}{fdecc8}
\definecolor{Initial}{HTML}{ffe2dd}
\definecolor{Human}{HTML}{fadec9}
\definecolor{AI}{HTML}{d3e5ef}
\definecolor{Direct}{HTML}{eee0da}
\definecolor{Indirect}{HTML}{dbeddb}
\definecolor{Visual Brush Selection}{HTML}{f5e0e9}
\definecolor{Widget Selection}{HTML}{fdecc8}
\definecolor{Inline Annotation}{HTML}{dbeddb}
\definecolor{Interactive Visual Click}{HTML}{ffe2dd}
\definecolor{Flowchart Manipulation}{HTML}{e8deee}
\definecolor{Tree Manipulation}{HTML}{fadec9}
\definecolor{Text Brush Selection}{HTML}{d3e5ef}
\definecolor{Text Input or Editing}{HTML}{ffffcc}
\definecolor{Inline Highlighting}{HTML}{e9fbe0}
\definecolor{Viewpoint Navigation}{HTML}{e8deee}
\definecolor{Sketch}{HTML}{c8e3f3}
\definecolor{Spreadsheet Manipulation}{HTML}{f1f0ef}
\definecolor{Speech}{HTML}{f1f0ef}
\definecolor{Global Software Manipulation}{HTML}{f1f0ef}
\definecolor{Drag and Drop}{HTML}{f1f0ef}

\definecolor{tablerowcolor}{rgb}{0.667,0.667,0.667 }
\definecolor{tablerowcolor2}{rgb}{0,0,0}
\definecolor{visual}{HTML}{e8efd9}
\definecolor{motion}{HTML}{fde7d5}
\definecolor{narrative}{HTML}{e2dce9}
\definecolor{audio}{HTML}{d6ebf2}
\definecolor{bluecrayola}{rgb}{0.12,0.46,1.0}

\begin{document}
\title[Prompting Generative AI with Interaction-Augmented Instructions]{Prompting Generative AI with \\Interaction-Augmented Instructions}

\author{Leixian Shen}
\orcid{0000-0003-1084-4912}
\affiliation{%
  \institution{The Hong Kong University of Science and Technology}
  \city{Hong Kong SAR}
  \country{China}
}
\email{lshenaj@connect.ust.hk}

\author{Haotian Li}
\orcid{0000-0001-9547-3449}
\affiliation{%
  \institution{Microsoft Research Asia}
  \city{Beijing}
  \country{China}
}
\email{haotian.li@microsoft.com}

\author{Yifang Wang}
\orcid{0000-0001-6267-9440}
\affiliation{%
  \institution{Northwestern University}
  \city{Evanston}
  \country{United States}
}
\email{yifang.wang@kellogg.northwestern.edu}

\author{Xing Xie}
\orcid{0000-0002-8608-8482}
\affiliation{%
  \institution{Microsoft Research Asia}
  \city{Beijing}
  \country{China}
}
\email{xingx@microsoft.com}

\author{Huamin Qu}
\orcid{0000-0002-3344-9694}
\affiliation{%
  \institution{The Hong Kong University of Science and Technology}
  \city{Hong Kong SAR}
  \country{China}
}
\email{huamin@cse.ust.hk}

\begin{abstract}
The emergence of generative AI (GenAI) models, including large language models and text-to-image models, has significantly advanced the synergy between humans and AI with not only their outstanding capability but more importantly, the intuitive communication method with text prompts. 
Though intuitive, text-based instructions suffer from natural languages' ambiguous and redundant nature. 
To address the issue, researchers have explored augmenting text-based instructions with interactions that facilitate precise and effective human intent expression, such as direct manipulation.
However, the design strategy of interaction-augmented instructions lacks systematic investigation, hindering our understanding and application.
To provide a panorama of interaction-augmented instructions, we propose a framework to analyze related tools from why, when, who, what, and how interactions are applied to augment text-based instructions. Notably, we identify four purposes for applying interactions, including restricting, expanding, organizing, and refining text instructions. The design paradigms for each purpose are also summarized to benefit future researchers and practitioners.

\end{abstract}



\maketitle

\section{Introduction}
Empowering humans with AI collaborators has been widely investigated and applied in various fields, such as decision making~\cite{Suh2023a}, mass communication~\cite{DVSurvey}, creative design~\cite{zhou2024understanding}, and scientific discovery~\cite{Liu2024b}.
In human-AI collaboration, one critical challenge is to convey human intent, such as the goal and mode of collaboration, effectively and efficiently to AI~\cite{Kim2024a}. 
Only by ensuring AI collaborators align with humans' intent can the collaboration be synergistic and productive~\cite{li2023ai}.
As a result, methods for conveying human intent to AI attract broad research interest, from direct interactions~\cite{Lehmann2024, Luera} to implicit understanding of human behavior~\cite{Shen2024, Gao2024}.

The recent emerging large-scale generative AI (GenAI) models, such as large language models (LLMs, \eg GPT-4o), and image or video generation models (\eg Sora), have considerably advanced this line of research.
Humans can instruct these models with natural language-based text prompts.
Though these instructions are intuitive and flexible, they also inherit the intrinsic disadvantages from natural languages, such as redundancy and ambiguity~\cite{NLISurvey}, which lowers the effectiveness and efficiency of human intent communication to AI collaborators.
To compensate for the disadvantage, researchers and practitioners have widely explored how to leverage novel interactions to enhance text-only prompts.
For example, one notable application of the design strategy is OpenAI Canvas~\cite{canvas}, which lets users specify the functional scope of text prompts to LLMs through brushing.
With such \textbf{interaction-augmented instructions}, human intent can be expressed in a more precise manner and thus AI collaborators can take more targeted actions.
Other examples of interaction-augmented instructions include visual brush selection on images, flowchart manipulation, inline text annotations or highlighting, \etc~
Though the strategy of enhancing text prompts with interactions has been widely applied, we notice that there still lacks systematic examination and summary of its design paradigms, hindering the re-use and innovation of related techniques.

\begin{figure*}[t]
  \centering
    \includegraphics[width=0.95\linewidth]{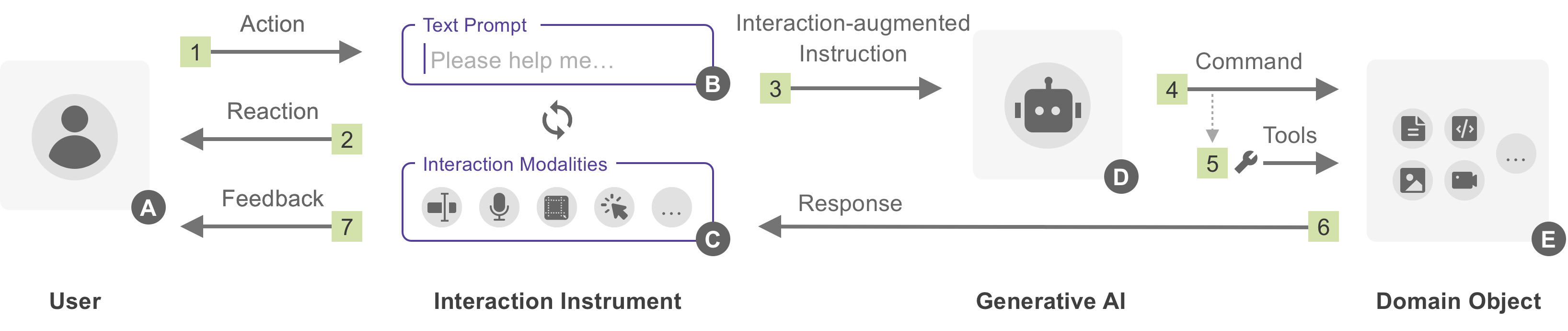}
     \vspace{-10px}
    \caption{GenAI-based instrumental interaction model.}
\label{fig:model}
  \vspace{-10px}
\end{figure*}

To fill the gap, we conducted a preliminary review of existing tools that support interaction-augmented instructions.
Built on the well-recognized instrumental interaction model~\cite{Beaudouin-Lafon2000}, we first formalize the role of interaction-augmented instructions in the communication between humans and GenAI with a GenAI-based instrumental interaction model.
The model helps us to understand \textit{when} interaction-augmented instructions are issued in the communication process, \textit{who} drives the formation of these instructions, and 
\textit{how} these instructions are applied on domain objects.
Next, we further investigate the motivation (\textit{why}) of applying interaction-augmented instructions, including restricting, expanding, organizing, and refining text instructions, and the detailed interaction types (\textit{what}) to convey human intent.
With the 4W1H framework, we analyze the existing tools that support interaction-augmented instructions and summarize the prominent design patterns (Appendix Tab.~\ref{tab: data}).
Based on the insights, we identify a series of research opportunities and hope to further enrich our knowledge of enhancing intent communication through interaction-augmented interaction from multiple aspects, such as further understanding human agency and technical implementation behind these paradigms.

\section{Related Work}

The rapid advancement of AI has not only facilitated the automation of numerous tasks but has also catalyzed human-AI communication and collaboration~\cite{Teevan2022, Wang2024h, Mueller2023, Chen2023f, Yildirim2023}. 
For example, 
Capel~\etal~\cite{Capel2023} presented a research landscape of human-centered AI.
Shen~\etal~\cite{Shen2024} reviewed over 400 papers across multiple domains and proposed the bidirectional human-AI alignment framework.
Rezwana and Maher~\cite{Rezwana2023} developed the Co-Creative Framework for Interaction design (COFI) and used the framework to categorize the interaction models in 92 existing human-AI co-creative systems.
Moruzzi and Margarido~\cite{Moruzzi2024} introduced a user-centered framework for human-AI co-creativity, defining key dimensions to understand the balance between user and system agency.
However, these frameworks are coarse-grained and primarily emphasize high-level theoretical human-AI collaboration.

To achieve seamless human-AI collaboration, effective communication between humans and AI is essential.
When communicating with GenAI, text prompts are the most commonly used approach.
Due to the ambiguous and redundant nature of natural languages, prior studies have shown that crafting precise and effective prompts impose high requirements on humans' cognitive capabilities~\cite{Zamfirescu-Pereira2023, Bansal, Subramonyam2023, Tankelevitch2024}.
To address the challenge, many studies incorporate interactions for creating effective GenAI prompts with minimum efforts (\eg \cite{canvas, Vaithilingam2024}).
To summarize potential interaction designs, Lehmann~\etal~\cite{Lehmann2024} investigated how different UIs influence users’ access to the functional capabilities of GenAI models. 
Luera~\etal~\cite{Luera} surveyed interface designs and patterns in GenAI applications, focusing on user-guided input modalities.
Gao~\etal~\cite{Gao2024} proposed a high-level taxonomy of human-LLM communication modes.
However, these frameworks focus primarily on summarizing what interactions are applied to enhance human-GenAI communication.
Our work aims to dig deeper to characterize the interplay between those interactions and text prompts with a framework that focuses on \textit{why} and \textit{what} interactions are applied to enhance text prompts, and \textit{who} drives the formation of interaction-augmented instructions.
The framework also includes \textit{when} these instructions were issued and \textit{how} they function.
Compared to prior research,
our fine-grained framework can have stronger descriptive power~\cite{beaudouin2004designing} to summarize the design paradigms of existing tools and more generative power to 
point out research opportunities (Sec.~\ref{sec:pattern}).

\begin{figure*}[t]
  \centering
    \includegraphics[width=\linewidth]{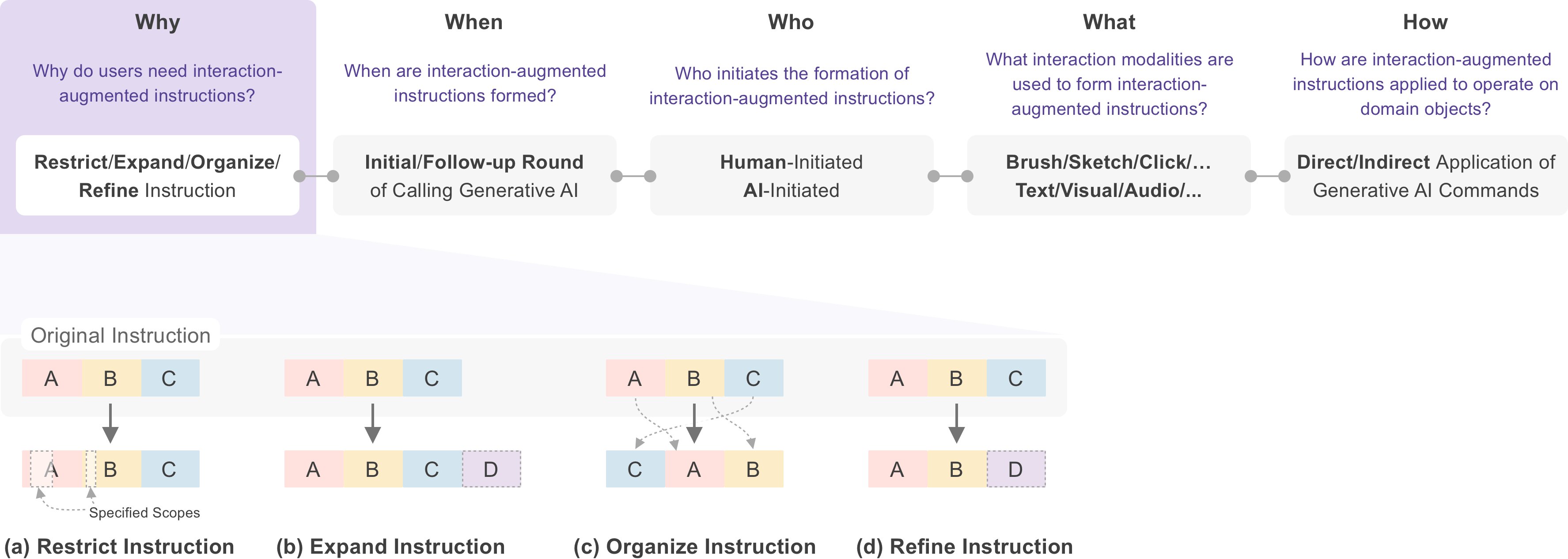}
     \vspace{-10px}
    \caption{The 4W1H analytical framework of existing GenAI tools featuring interaction-augmented instructions. The ``\textit{Why}'' dimension describes four impact types of interactions on instructions.}
\label{fig:analysis}
  \vspace{-10px}
\end{figure*}

\section{GenAI-based Instrumental Interaction Model}
To understand the design paradigms of interaction-augmented instructions, we need to formalize the communication process between humans and GenAI.
During the process, GenAI interprets humans' intention expressed in natural language and responds accordingly, either by taking direct actions or invoking external tools to operate on specific instruction objects when necessary.

Inspired by the report edited by Butler~\etal~\cite{Teevan2022}, we realize that the interplay between humans, interaction instruments, GenAI, and target domain objects forms a GenAI-based version of the well-established \textit{Instrumental Interaction Model}~\cite{Beaudouin-Lafon2000}, where interaction instruments serve as mediators between users and domain objects.
However, the original model relies on specialized instruments (\eg scrolling bar) which often provide a clear mapping from user actions to operations on target objects.
The model is hard to fully capture the usage of GenAI for flexible interactions between humans and domain objects.
With the participation of GenAI, users can leverage free-form text prompts as primary instruments to issue instructions with diverse intent, and GenAI will translate these instructions to executable operations on domain objects. 
Though flexible, these text prompts also suffer from the inherent ambiguity of natural language, 
To address the issue, researchers further explore enhancing text prompts with additional interactions, such as highlighting text segment with text brushing selection~\cite{canvas}, to achieve more precise operations on domain objects.
To describe the process, we propose a new \textit{GenAI-based Instrumental Interaction Model} in Fig.~\ref{fig:model}.

In this model, interaction instruments serve as the method for human (A)-GenAI (D) communication to achieve operations on domain objects (E) based on user intent.
Users act with instruments (1) via text prompts (B) and other interaction modalities (C), such as through text brushing, canvas clicking, widget selection, and textual inline annotation (all summarized interaction modalities are shown in Appendix Tab.~\ref{tab: data}), which collectively transform users' intents into interaction-augmented instructions for GenAI (3). 
GenAI, in turn, directly operates on domain objects (4), \eg generated text or images within GenAI's ability or produces commands to trigger other tools to operate on domain objects (5).
Instruments also provide reactions (2) that allow users to refine their actions and provide feedback (7) as the instructions are executed by GenAI and applied to domain objects (6).
In addition, the process is iterative.

\section{Framework}\label{sec:framework}
With a model to characterize the role of interaction-augmented instructions in human-GenAI communication, we propose a framework based on 4W1H—Why, When, Who, What, and How—to analyze existing interaction-augmented instruction paradigms (Fig.~\ref{fig:analysis}).

\textit{\textbf{Why} do users need interaction-augmented instructions?}
The impact of interaction-augmented instructions on domain objects varies based on user intents~\cite{Rezwana2023, Wang2024h, Subramonyam2023}. 
We categorized the impact of user interactions on original text prompts into four types (Fig.~\ref{fig:analysis}): 
(a) \textit{Restrict Instruction}: interactions are applied to restrict the scope of AI actions (\eg a specific part of domain objects) when text prompts are vague;
(b) \textit{Expand Instruction}: interactions are applied to offer more guidance when text prompts provide limited information;
(c) \textit{Organize Instruction}: interactions are applied  to structure information for AI's better understanding of free-form text prompts;
(d) \textit{Refine Instruction}: interactions are applied to improve the clarity of ambiguous text prompts.

\textit{\textbf{When} are interaction-augmented instructions formed?}
Interaction-augmented instructions can be formulated either in the \textit{initial} or \textit{follow-up} round of calling GenAI~\cite{Bansal, Shen2024}.
In the initial round, users carefully prepare instructions based on application requirements before passing them to GenAI, \ie Fig.~\ref{fig:model} (1 and 3).
In the follow-up rounds, users make further decisions or improvements after receiving GenAI execution feedback, \ie Fig.~\ref{fig:model} (6, 2 and 7).

\textit{\textbf{Who} initiates the formation of interaction-augmented instructions?}
The formulation of interactive prompts can be either \textit{human-initiated} or \textit{AI-initiated}~\cite{Capel2023, Subramonyam2023, Tankelevitch2024}. 
In human-initiated mode (Fig.~\ref{fig:model}-A), users craft high-quality instructions to ensure GenAI understand the intent or refine the instructions based on GenAI decisions. 
In AI-initiated mode (Fig.~\ref{fig:model}-D), AI assists users by suggesting ways to specify better instructions or providing recommendations for improvement, followed by humans' final decision based on these suggestions, thus improving AI's explainability.

\textit{\textbf{What} interaction modalities are used to form interaction-augmented instructions?}
Domain objects (Fig.~\ref{fig:model}-E) in applications vary widely (\eg text, image, code, canvas, \etc) and can all be used to specify instructions~\cite{Luera}. 
Users can employ various actions (\eg clicking, selecting, brushing, sketching, typing, \etc) to generate additional interactions that augment the original text prompts (Fig.~\ref{fig:model}-B).

\textit{\textbf{How} are interaction-augmented instructions applied to operate on domain objects?}
The commands provided by GenAI based on interaction-augmented instructions can be \textit{directly} or \textit{indirectly} applied on domain objects.
In direct mode, GenAI directly understands and executes commands to operate on domain objects (Fig.~\ref{fig:model}-4), \eg in writing and image editing tasks.
In indirect mode, GenAI understands the instruction and calls other tools to operate on domain objects (Fig.~\ref{fig:model}-D), \eg utilizing a tool to generate animation effects on specific graphic elements.



\section{Corpus Collection and Analysis}
We applied the framework in Sec.~\ref{sec:framework} to analyze the design of existing tools with interaction-augmented instructions.
The tools under analysis must adhere to the following criteria: (1) the system must be GenAI-based; (2) interaction with GenAI must involve text prompts; (3) apart from text prompts, at least one supplementary form of interaction is used to augment the text prompts.
Our data collection process commenced by sourcing papers from key HCI conferences: CHI, UIST, CSCW, and IUI, using targeted keyword searches (\eg ``GenAI'', ``Generative AI'', ``LLM'', ``Large Language Model'', \etc) starting from November 2022 (following the release of OpenAI's ChatGPT). Additionally, we curated a set of recent survey or position papers on human-GenAI interaction as seed papers~\cite{Luera, Subramonyam2023, Shen2024, Gao2024, Zhang2024a, Li2023c}. Subsequently, we analyzed the corpus of the seed papers along with their references to amplify our dataset.
Finally, we collected 52 tools
(see Appendix Tab.~\ref{tab: data}) 
with interaction-augmented instructions, covering various tasks like coding, writing, visualization authoring, image generation and editing, \etc~

\section{Design Paradigms}\label{sec:pattern}
We will discuss the identified patterns using our framework. 
These patterns can guide future human-GenAI interaction design and inspire the exploration of alternative paradigms. 
The discussion is organized primarily by the ``Why'' dimension, with further subdivision into four categories based on ``When'' and ``Who'': human-initiated instruction preparation, AI-initiated instruction preparation, human-initiated post-invocation enhancement, and AI-initiated post-invocation enhancement. 
Each category is analyzed with respect to corresponding ``What'' and ``How''.

\subsection{Restrict Instruction}
User text prompts are frequently vague when defining domain-specific objects of interest.
Interactions allow users to restrict the scope of AI actions, aligning outputs more closely with human intents (29/52)\footnote{This format indicates that 29 tools out of 52 are under this category.}.

\begin{figure*}[t]
  \centering
    \includegraphics[width=\linewidth]{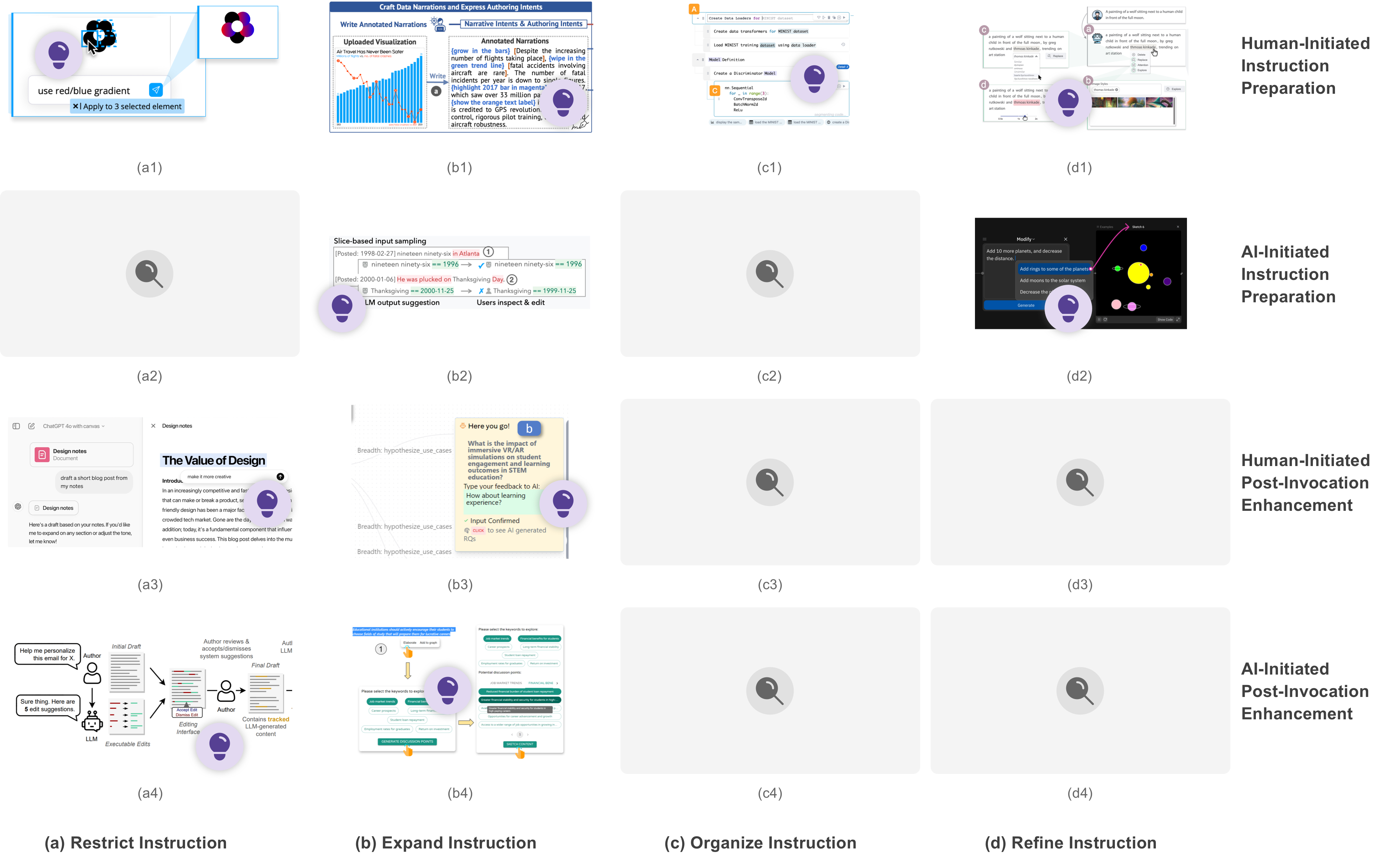}
     \vspace{-10px}
    \caption{Examples of interaction-augmented instructions, displayed by four impact types of interaction on instructions (``Why'') and four sub-categories based on ``When'' and ``Who'' in Sec.~\ref{sec:pattern}. 
    In each example, \includegraphics[width=0.017\linewidth]{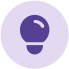}
    denotes the core interaction design, while \includegraphics[width=0.017\linewidth]{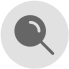}
    indicates the absence of typical paradigm instances identified in our paper, leaving for future exploration.
    The examples are from (a1)~\cite{Masson2023b}, (a3)~\cite{canvas}, (a4)~\cite{Laban2023}, (b1)~\cite{dataplaywright}, (b2)~\cite{Wu2023c}, (b3)~\cite{Liu2024b}, (b4)~\cite{Zhang2023b}, (c1)~\cite{Zhu}, (d1)~\cite{Wang2024g}, and (d2)~\cite{Angert2023}.
    }
\label{fig:example}
  \vspace{-1em}
\end{figure*}

\bpstart{Human-Initiated Instruction Preparation (14/29)}
These paradigms focus on instruction setup before invoking GenAI within a single interaction round. Users actively leverage diverse interaction modalities to restrict the scope of AI actions, typically interacting on user-uploaded domain objects.
Interactive visual click selection is a widely adopted approach, enabling users to directly select target objects via mouse clicks (Fig.~\ref{fig:example}-a1), such as SVG elements~\cite{Masson2023b}, graph structure nodes~\cite{Zhang2024,Xie2024a}, and segmented image regions~\cite{Liu2023c}, which allows explicit processing by GenAI, and UI components on mobile canvas~\cite{Gao2024a}, which implicitly execute GenAI' decisions on the canvas.  
Similarly, several tools augment text prompts with preconfigured widget options, enabling users to restrict action scopes~\cite{Brade2023, Wang2023a}. Furthermore, DirectGPT~\cite{Masson2023b} supports dragging visual elements and dropping them into text prompts.  
Visual brush selection is another common interaction method, used for controllable image generation~\cite{Chung2023}, region-specific image editing~\cite{Liu2024d}, visual question answering~\cite{Choe2024}, and implicit video editing~\cite{Tilekbay}.  
In textual domains (\eg writing and coding), inline textual highlighting~\cite{Jung2023} and text brush selection~\cite{Reza2024, Fok2024} are effective ways to restrict AI action scope. The selection of the text brush often triggers further widget configurations~\cite{Reza2024, Fok2024}.

\bpstart{Human-Initiated Post-Invocation Enhancement (4/29)}
In contrast to human-initiated instruction preparation paradigms, human-initiated post-invocation enhancement paradigms focus on multi-turn interactions, refining instructions after GenAI invocation while leveraging similar interaction modalities. 
In this context, the domain objects are outputs from a previous AI generation round, which users may wish to fine-tune or regenerate. Interactions can help restrict the scope for subsequent instructions~\cite{Aksan2018, canvas, Wang2024g, cursor}.  
For example, OpenAI Canvas~\cite{canvas} (Fig.~\ref{fig:example}-a3) allows users to select portions of generated text or code for further querying, while PromptCharm~\cite{Wang2024g} enables users to select the image of interest from the generated set for subsequent exploration.

\bpstart{AI-Initiated Post-Invocation Enhancement (12/29)}
In these paradigms, users usually provide coarse-grained instructions, to further confirm users' intent, AI can generate fine-grained options for users' selection. In this way, the action space of AI can be restricted.
The primary interaction modality is widget selection, which could be categorized into two types. 
One type is \textit{accepting user feedback to guide subsequent human-GenAI interactions}. 
For example, in the binary selection space, InkSync~\cite{Laban2023} (Fig.~\ref{fig:example}-a4) uses inline highlighting to show all LLM-proposed modifications to user writing and provides an accept/dismiss dialogue menu.
In the multi-class selection space, ConstitutionMaker~\cite{Petridis2024} guides users through iterative dialogue with multiple-choice questions to customize personal chatbots, this paradigm is also common in other conversational tools~\cite{Bursztyn2021, Baek2023, Gero2022, Chen2024d}. 
In the unknown selection space, Luminate~\cite{Suh2023} uses an initial LLM to generate task-related dimensions and widget configurations, enabling users to specify their intents, followed by another LLM to fulfill the requirements.
Unlike the above explicit human-GenAI interaction, LangAware~\cite{Chen2023e} offers fine-grained options for natural language interaction on mobile devices. Users confirm these options and implicitly execute them on connected sensors after LLM decision-making.
The other type is \textit{providing customization options for final outputs without additional GenAI invocation}~\cite{Andrews2024, Vaithilingam2024, Tang2024}. 
For example, DynaVis~\cite{Vaithilingam2024} dynamically generates rich widgets for users to configure visualizations. Alternatively, AI performs most tasks, leaving users with simple review and confirmation, such as PDFChatAnnotator~\cite{Tang2024}, which extracts user-specified PDF content as cards for review.

\subsection{Expand Instruction}  
For complex tasks, user text prompts may lack sufficient detail or breadth. User interactions can expand instructions to guide GenAI more effectively (19/52).  

\bpstart{Human-Initiated Instruction Preparation (9/19)}  
In these paradigms, users expand initial text prompts through interactions to provide GenAI with richer context or clearer instructions for complex tasks. The interaction modalities in this category are often more complex and task-specific. 
For instance, Tableau AI~\cite{tableauai} integrates a chatbot with Tableau’s traditional WIMP interface, where all real-time interactions with visualizations provide contextual input for the chatbot. 
G-VOILA~\cite{Wang2024i} and XR-Objects~\cite{Dogan2024} enrich text queries with environmental context through viewpoint navigation using augmented reality glasses. 
Tools like ChatScratch~\cite{Chen2024d} and DrawTalking~\cite{Rosenberg} use sketches to augment text prompts for creative drawing tasks. 
FigurA11y~\cite{Singh2024} incorporates a set of selectable guidelines for generating figure alt text.  
For more complex outputs (\eg video), users often rely on extended instructions with implicit integration of auxiliary tools. For example, NotePlayer~\cite{NotePlayer} enhances the context for tutorial video generation by allowing inline annotations in notebooks. 
LAVE~\cite{Wang2024} enables users to select a series of relevant videos as input material for text-driven video editing.
Data Playwright~\cite{dataplaywright} (Fig.~\ref{fig:example}-b1) encodes data video authoring commands as inline annotations within narrations, enriching the context of both narrative and authoring intents.

\bpstart{AI-Initiated Instruction Preparation (1/19)}  
When users struggle to extend instructions—especially due to unfamiliarity with task-specific options or the overwhelming number of input candidates—AI-initiated instruction preparation paradigms can effectively recommend instruction candidates (\eg examples in text prompts), streamlining the preparation process.  
For instance, ScatterShot~\cite{Wu2023c} (Fig.~\ref{fig:example}-b2) builds high-quality demonstration sets for LLMs by iteratively segmenting unlabeled user data into task-specific patterns for human review and selection. 

\bpstart{Human-Initiated Post-Invocation Enhancement (5/19)}  
Exploratory tasks often require multiple rounds of interaction with GenAI to achieve the desired outcome. 
Flowchart manipulation is a common interaction modality that helps users organize their thoughts and iteratively refine instructions~\cite{Liu2024b, Kim2023c, Suh2023a, Angert2023}.  
For example, CoQuest~\cite{Liu2024b} (Fig.~\ref{fig:example}-b3) facilitates LLM-based research question exploration by generating a series of idea nodes from high-level text prompts. Users can explore each node in depth, providing feedback to guide subsequent iterations.
Sensecape~\cite{Suh2023a} enables users to extract key points from LLM-generated text outputs and automatically organize them into multi-level representations, facilitating navigation and exploration across complex information spaces.

\bpstart{AI-Initiated Post-Invocation Enhancement (4/19)}  
In exploratory tasks, some tools utilize AI to guide users in expanding subsequent instructions. 
They leverage AI to break down tasks, provide structure, and suggest further actions for enhanced task execution. 
For instance, ExploreLLM~\cite{Ma2024} automatically decomposes user text prompts into sub-tasks, creating nodes that represent each sub-task, and allowing users to specify personal contexts and preferences for further exploration. 
Similarly, VISAR~\cite{Zhang2023b} (Fig.~\ref{fig:example}-b4) automatically recommends possible key aspects of writing for users to select and further explore, while GAM~\cite{Liu2023e} recommends data analysis utterance in spreadsheets.
Graphologue~\cite{Jiang2023} generates inline annotated LLM responses and parses them into diagrams to facilitate user exploration and decision-making.

\subsection{Organize Instruction}  
User-provided text prompts are often numerous and unstructured, hindering effective task and instruction interpretation. Structuring these prompts into organized formats can enhance task comprehension and execution (6/52).

\bpstart{Human-Initiated Instruction Preparation (6/6)}  
Paradigms in this group focus on user-driven efforts to organize instructions before GenAI invocation. 
Among these formats, tree structures are widely used to represent hierarchical relationships and logical flows~\cite{Feng2024, Zhu, Zhang2023b, Wu2022e}.  
For example, CoPrompt~\cite{Feng2024} introduces a tree-based representation to organize shared coding prompts among collaborators, supporting referencing, requesting, sharing, and linking prompts seamlessly. Similarly, CoLadder~\cite{Zhu} (Fig.~\ref{fig:example}-c1) applies a tree structure to organize hierarchical coding tasks. VISAR~\cite{Zhang2023b} adopts a tree structure to aid writers in brainstorming and revising hierarchical writing goals, streamlining the creative process.  
Low-code LLM~\cite{Cai2024} utilizes flowcharts to help users plan complex tasks step by step, ensuring clarity in task execution. 
Additionally, Kim et al.~\cite{Kim2023d} propose a modular framework comprising cells, generators, and lenses, allowing users to organize LLM applications flexibly. It  supports various widgets for configuring parameters.

\subsection{Refine Instruction}  
User text prompts are often imprecise, resulting in suboptimal model outputs and necessitating additional rounds of interaction. Users can clarify their instructions by incorporating interactive refinement mechanisms, improving GenAI accuracy and reducing unnecessary iterations (3/52).

\bpstart{Human-Initiated Instruction Preparation (2/3)}  
In this category, users actively explore and refine their prompts to ensure better alignment between user intent and model understanding. 
For instance, PromptCharm~\cite{Wang2024g} (Fig.~\ref{fig:example}-d1) enables users to refine text-to-image prompts by exploring a large database of styles. Users can search, compare, and modify styles through interactive lists. 
DreamSheets~\cite{Almeda2024} leverages spreadsheet-based manipulation to quickly enumerate different prompt variations for text-to-image generation, enabling users to generate all variations simultaneously and efficiently explore the generated space.

\bpstart{AI-Initiated Instruction Preparation (2/3)}  
Paradigms in this group involve tools proactively guiding users by suggesting optimal prompts. For instance, autocomplete features during text input~\cite{Angert2023} (Fig.~\ref{fig:example}-d2) assist users by completing partial phrases or suggesting contextually relevant continuations. Additionally, candidate recommendations~\cite{Wang2024g} present users with optimized options after initial input, helping them construct more precise and effective prompts. 


\section{Conclusion and Future Work}
This paper provides a preliminary analysis and summary of the design paradigms for interaction-augmented instructions in existing GenAI tools, offering guidance for future human-GenAI interaction design and inspiring the exploration of additional paradigms. In the future, we plan to further investigate the following directions. 

\nistart{Explore More Paradigms for Interaction-Augmented Instructions}
Our framework offers a starting point to explore interaction-augmented instructions. Fig.~\ref{fig:example} presents an initial two-dimensional design space and corresponding examples. 
Several subspaces remain unexplored, offering promising directions for future investigation. 
However, it is also worth considering whether some subspaces are better suited for purely text-based prompts or traditional AI models. 
Theoretically, any interaction-enhanced instruction could be replicated by optimizing text prompts. Therefore, carefully evaluating the trade-offs between text-only prompts and interaction-augmented instructions is essential~\cite{Tankelevitch2024}. 
For example, when specifying domain objects, interactive selection methods like clicking, brushing, or lassoing on a cluttered canvas can be more intuitive, accurate, and efficient than natural language~\cite{WonderFlow, lin2023inksight}. However, in selections with semantic features or between overlapping elements, natural language can offer greater efficiency and clarity~\cite{dataplaywright, dataplayer}.

\nistart{Build GenAI Applications with Interaction-Augmented Instructions}
Our ``Why'' dimension highlights four key challenges users often face when crafting text prompts: overly broad scopes, limited guidance, disorganized structures, and imprecise expressions. Correspondingly, we have summarized how interaction can help restrict, expand, organize, and refine original text prompts. 
When designing GenAI applications, our framework and identified patterns can guide decisions about whether to incorporate interaction-augmented instructions and select appropriate interaction paradigms~\cite{Tankelevitch2024}. Moreover, effective GenAI applications can integrate multiple paradigms across different stages of interaction to address diverse user needs.

\nistart{Understand Human Agency and Technical Implementation Behind Interaction-Augmented Instructions}
We observe that different paradigms of interaction-augmented instructions involve different technical implementation to achieve varying levels of human agency and AI automation~\cite{Heer2019}. 
Through analysis with our framework, we identify different interaction granularities and mediums, reflecting variations in human involvement and influencing the underlying AI capabilities required.  
For example, indirect operations on domain objects require AI to select and apply additional tools.
It elevates the automation level but may diminish users' agency.
To make these decisions, advanced planning and tool application capabilities can be necessary for these AI models. 
While our current study focuses on interactions, future research is necessary to explore the interplay among interaction, human agency, and technical implementation, offering systematic guidance and further directions for enhancing human-GenAI interaction.



\bibliographystyle{ACM-Reference-Format}

\bibliography{main}

\clearpage
\appendix
\arrayrulecolor[gray]{0.8}
\setlength{\fboxsep}{0.5pt}
\begin{table*}[t]
\footnotesize
\centering
\caption{Overview of existing tools featuring interaction-augmented instructions, labeled within our analysis framework (Sec.~\ref{sec:framework}). The tools are sorted by ``Why'', ``When'', ``Who'', ``How'', and ``What''.}
\vspace{-10px}
\label{tab: data}
\renewcommand\arraystretch{0.84}
\setlength{\tabcolsep}{2.9mm}{
\begin{tabular}{lccllllc}
\toprule
\textbf{Name} & \textbf{Source} & \textbf{Year} & \textbf{Why} & \textbf{When} & \textbf{Who} & \textbf{What} & \textbf{How} \\
\midrule

Choe et al.~\cite{Choe2024} & TVCG & 2024 & \transparentcolorbox{Restrict}{Restrict} & \transparentcolorbox{Initial}{Initial} & \transparentcolorbox{Human}{Human} & \transparentcolorbox{Visual Brush Selection}{Visual Brush Selection} & \transparentcolorbox{Direct}{Direct} \\ \hline
MagicQuill~\cite{Liu2024d} & arXiv & 2024 & \transparentcolorbox{Restrict}{Restrict} & \transparentcolorbox{Initial}{Initial} & \transparentcolorbox{Human}{Human} & \transparentcolorbox{Visual Brush Selection}{Visual Brush Selection} & \transparentcolorbox{Direct}{Direct} \\ \hline
PromptPaint~\cite{Chung2023} & UIST & 2023 & \transparentcolorbox{Restrict}{Restrict} & \transparentcolorbox{Initial}{Initial} & \transparentcolorbox{Human}{Human} & \transparentcolorbox{Visual Brush Selection}{Visual Brush Selection} & \transparentcolorbox{Direct}{Direct} \\ \hline
Data Formulator~\cite{Wang2023a} & TVCG & 2024 & \transparentcolorbox{Restrict}{Restrict} & \transparentcolorbox{Initial}{Initial} & \transparentcolorbox{Human}{Human} & \transparentcolorbox{Widget Selection}{Widget Selection} & \transparentcolorbox{Direct}{Direct} \\ \hline
CML~\cite{Jung2023} & IUI & 2023 & \transparentcolorbox{Restrict}{Restrict} & \transparentcolorbox{Initial}{Initial} & \transparentcolorbox{Human}{Human} &  \transparentcolorbox{Inline Highlighting}{Inline Highlighting} & \transparentcolorbox{Direct}{Direct} \\ \hline
DirectGPT~\cite{Masson2023b} & CHI & 2024 & \transparentcolorbox{Restrict}{Restrict} & \transparentcolorbox{Initial}{Initial} & \transparentcolorbox{Human}{Human} & \begin{tabular}[c]{@{}l@{}} \transparentcolorbox{Interactive Visual Click}{Interactive Visual Click}\\   \transparentcolorbox{Drag and Drop}{Drag and Drop}\end{tabular} & \transparentcolorbox{Direct}{Direct} \\ \hline
Mindalogue~\cite{Zhang2024} & arXiv & 2024 & \transparentcolorbox{Restrict}{Restrict} & \transparentcolorbox{Initial}{Initial} & \transparentcolorbox{Human}{Human} &  \transparentcolorbox{Interactive Visual Click}{Interactive Visual Click} & \transparentcolorbox{Direct}{Direct} \\ \hline
WaitGPT~\cite{Xie2024a} & UIST & 2024 & \transparentcolorbox{Restrict}{Restrict} & \transparentcolorbox{Initial}{Initial} & \transparentcolorbox{Human}{Human} &  \transparentcolorbox{Interactive Visual Click}{Interactive Visual Click} & \transparentcolorbox{Direct}{Direct} \\ \hline
InternGPT~\cite{Liu2023c} & arXiv & 2023 & \transparentcolorbox{Restrict}{Restrict} & \transparentcolorbox{Initial}{Initial} & \transparentcolorbox{Human}{Human} &  \transparentcolorbox{Interactive Visual Click}{Interactive Visual Click} & \transparentcolorbox{Direct}{Direct} \\ \hline
Promptify~\cite{Brade2023} & UIST & 2023 & \begin{tabular}[c]{@{}l@{}}\transparentcolorbox{Restrict}{Restrict}\\  \transparentcolorbox{Expand}{Expand}\end{tabular} & \begin{tabular}[c]{@{}l@{}}\transparentcolorbox{Initial}{Initial}\\  \transparentcolorbox{Follow-up}{Follow-up}\end{tabular} & \transparentcolorbox{Human}{Human} & \begin{tabular}[c]{@{}l@{}} \transparentcolorbox{Interactive Visual Click}{Interactive Visual Click}\\  \transparentcolorbox{Widget Selection}{Widget Selection}\end{tabular} & \transparentcolorbox{Direct}{Direct} \\ \hline
ABScribe~\cite{Reza2024} & CHI & 2024 & \transparentcolorbox{Restrict}{Restrict} & \transparentcolorbox{Initial}{Initial} & \transparentcolorbox{Human}{Human} & \begin{tabular}[c]{@{}l@{}} \transparentcolorbox{Text Brush Selection}{Text Brush Selection}\\  \transparentcolorbox{Widget Selection}{Widget Selection}\end{tabular} & \transparentcolorbox{Direct}{Direct} \\ \hline
Qlarify~\cite{Fok2024} & UIST & 2024 & \transparentcolorbox{Restrict}{Restrict} & \transparentcolorbox{Initial}{Initial} & \transparentcolorbox{Human}{Human} & \begin{tabular}[c]{@{}l@{}} \transparentcolorbox{Text Brush Selection}{Text Brush Selection}\\  \transparentcolorbox{Widget Selection}{Widget Selection}\end{tabular} & \transparentcolorbox{Direct}{Direct} \\ \hline
ExpressEdit~\cite{Tilekbay} & IUI & 2024 & \transparentcolorbox{Restrict}{Restrict} & \transparentcolorbox{Initial}{Initial} & \transparentcolorbox{Human}{Human} & \transparentcolorbox{Visual Brush Selection}{Visual Brush Selection} & \transparentcolorbox{Indirect}{Indirect} \\ \hline
EasyAsk~\cite{Gao2024a} & IMWUT & 2024 & \transparentcolorbox{Restrict}{Restrict} & \transparentcolorbox{Initial}{Initial} & \transparentcolorbox{Human}{Human} &  \transparentcolorbox{Interactive Visual Click}{Interactive Visual Click} & \transparentcolorbox{Indirect}{Indirect} \\ \hline
DeepWriting~\cite{Aksan2018} & CHI & 2018 & \transparentcolorbox{Restrict}{Restrict} & \transparentcolorbox{Follow-up}{Follow-up} & \transparentcolorbox{Human}{Human} & \transparentcolorbox{Widget Selection}{Widget Selection} & \transparentcolorbox{Direct}{Direct} \\ \hline
CURSOR~\cite{cursor} & Industry & 2023 & \transparentcolorbox{Restrict}{Restrict} & \transparentcolorbox{Follow-up}{Follow-up} & \transparentcolorbox{Human}{Human} &  \transparentcolorbox{Text Brush Selection}{Text Brush Selection} & \transparentcolorbox{Direct}{Direct} \\ \hline
OpenAI Canvas~\cite{canvas} & Industry & 2024 & \transparentcolorbox{Restrict}{Restrict} & \transparentcolorbox{Follow-up}{Follow-up} & \begin{tabular}[c]{@{}l@{}}\transparentcolorbox{Human}{Human}\\  \transparentcolorbox{AI}{AI}\end{tabular} & \begin{tabular}[c]{@{}l@{}} \transparentcolorbox{Text Brush Selection}{Text Brush Selection}\\  \transparentcolorbox{Widget Selection}{Widget Selection}\end{tabular} & \transparentcolorbox{Direct}{Direct} \\ \hline
DynaVis~\cite{Vaithilingam2024} & CHI & 2024 & \transparentcolorbox{Restrict}{Restrict} & \transparentcolorbox{Follow-up}{Follow-up} & \transparentcolorbox{AI}{AI} & \transparentcolorbox{Widget Selection}{Widget Selection} & \transparentcolorbox{Direct}{Direct} \\ \hline
ConstitutionMaker~\cite{Petridis2024} & IUI & 2024 & \transparentcolorbox{Restrict}{Restrict} & \transparentcolorbox{Follow-up}{Follow-up} & \transparentcolorbox{AI}{AI} & \transparentcolorbox{Widget Selection}{Widget Selection} & \transparentcolorbox{Direct}{Direct} \\ \hline
Luminate~\cite{Suh2023} & CHI & 2024 & \transparentcolorbox{Restrict}{Restrict} & \transparentcolorbox{Follow-up}{Follow-up} & \transparentcolorbox{AI}{AI} & \transparentcolorbox{Widget Selection}{Widget Selection} & \transparentcolorbox{Direct}{Direct} \\ \hline
PDFChatAnnotator~\cite{Tang2024} & IUI & 2024 & \transparentcolorbox{Restrict}{Restrict} & \transparentcolorbox{Follow-up}{Follow-up} & \transparentcolorbox{AI}{AI} & \transparentcolorbox{Widget Selection}{Widget Selection} & \transparentcolorbox{Direct}{Direct} \\ \hline
PromptCrafter~\cite{Baek2023} & ICML WS & 2023 & \transparentcolorbox{Restrict}{Restrict} & \transparentcolorbox{Follow-up}{Follow-up} & \transparentcolorbox{AI}{AI} & \transparentcolorbox{Widget Selection}{Widget Selection} & \transparentcolorbox{Direct}{Direct} \\ \hline
Sparks~\cite{Gero2022} & DIS & 2022 & \transparentcolorbox{Restrict}{Restrict} & \transparentcolorbox{Follow-up}{Follow-up} & \transparentcolorbox{AI}{AI} & \transparentcolorbox{Widget Selection}{Widget Selection} & \transparentcolorbox{Direct}{Direct} \\ \hline
Bursztyn et al.~\cite{Bursztyn2021} & CHI EA & 2021 & \transparentcolorbox{Restrict}{Restrict} & \transparentcolorbox{Follow-up}{Follow-up} & \transparentcolorbox{AI}{AI} & \transparentcolorbox{Widget Selection}{Widget Selection} & \transparentcolorbox{Direct}{Direct} \\ \hline
InkSync~\cite{Laban2023} & UIST & 2024 & \transparentcolorbox{Restrict}{Restrict} & \transparentcolorbox{Follow-up}{Follow-up} & \transparentcolorbox{AI}{AI} & \begin{tabular}[c]{@{}l@{}} \transparentcolorbox{Widget Selection}{Widget Selection}\\   \transparentcolorbox{Inline Highlighting}{Inline Highlighting}\end{tabular} & \transparentcolorbox{Direct}{Direct} \\ \hline
AiCommentator~\cite{Andrews2024} & IUI & 2024 & \transparentcolorbox{Restrict}{Restrict} & \transparentcolorbox{Follow-up}{Follow-up} & \transparentcolorbox{AI}{AI} & \transparentcolorbox{Widget Selection}{Widget Selection} & \transparentcolorbox{Indirect}{Indirect} \\ \hline
LangAware~\cite{Chen2023e} & UIST & 2023 & \transparentcolorbox{Restrict}{Restrict} & \transparentcolorbox{Follow-up}{Follow-up} & \transparentcolorbox{AI}{AI} & \transparentcolorbox{Widget Selection}{Widget Selection} & \transparentcolorbox{Indirect}{Indirect} \\ \hline
Tableau AI~\cite{tableauai} & Industry & 2024 & \transparentcolorbox{Expand}{Expand} & \transparentcolorbox{Initial}{Initial} & \transparentcolorbox{Human}{Human} &  \transparentcolorbox{Global Software Manipulation}{Software Manipulation} & \transparentcolorbox{Direct}{Direct} \\ \hline
G-VOILA~\cite{Wang2024i} & IMWUT & 2024 & \transparentcolorbox{Expand}{Expand} & \transparentcolorbox{Initial}{Initial} & \transparentcolorbox{Human}{Human} &  \transparentcolorbox{Viewpoint Navigation}{Viewpoint Navigation} & \transparentcolorbox{Direct}{Direct} \\ \hline
XR-Objects~\cite{Dogan2024} & UIST & 2024 & \transparentcolorbox{Expand}{Expand} & \transparentcolorbox{Initial}{Initial} & \transparentcolorbox{Human}{Human} & \begin{tabular}[c]{@{}l@{}} \transparentcolorbox{Viewpoint Navigation}{Viewpoint Navigation}\\  \transparentcolorbox{Widget Selection}{Widget Selection}\end{tabular} & \transparentcolorbox{Direct}{Direct} \\ \hline
ChatScratch~\cite{Chen2024d} & CHI & 2024 & \begin{tabular}[c]{@{}l@{}}\transparentcolorbox{Expand}{Expand}\\  \transparentcolorbox{Restrict}{Restrict}\end{tabular} & \begin{tabular}[c]{@{}l@{}}\transparentcolorbox{Initial}{Initial}\\  \transparentcolorbox{Follow-up}{Follow-up}\end{tabular} & \begin{tabular}[c]{@{}l@{}}\transparentcolorbox{Human}{Human}\\  \transparentcolorbox{AI}{AI}\end{tabular} & \begin{tabular}[c]{@{}l@{}} \transparentcolorbox{Sketch}{Sketch}\\  \transparentcolorbox{Widget Selection}{Widget Selection}\end{tabular} & \transparentcolorbox{Direct}{Direct} \\ \hline
FigurA11y~\cite{Singh2024} & IUI & 2024 & \transparentcolorbox{Expand}{Expand} & \transparentcolorbox{Initial}{Initial} & \transparentcolorbox{Human}{Human} & \transparentcolorbox{Widget Selection}{Widget Selection} & \transparentcolorbox{Direct}{Direct} \\ \hline
DrawTalking~\cite{Rosenberg} & UIST & 2024 & \transparentcolorbox{Expand}{Expand} & \transparentcolorbox{Initial}{Initial} & \transparentcolorbox{Human}{Human} & \begin{tabular}[c]{@{}l@{}} \transparentcolorbox{Sketch}{Sketch}\\  \transparentcolorbox{Speech}{Speech}\end{tabular} & \transparentcolorbox{Indirect}{Indirect} \\ \hline
LAVE~\cite{Wang2024} & IUI & 2024 & \transparentcolorbox{Expand}{Expand} & \transparentcolorbox{Initial}{Initial} & \transparentcolorbox{Human}{Human} & \transparentcolorbox{Widget Selection}{Widget Selection} & \transparentcolorbox{Indirect}{Indirect} \\ \hline
Data Playwright~\cite{dataplaywright} & TVCG & 2024 & \transparentcolorbox{Expand}{Expand} & \transparentcolorbox{Initial}{Initial} & \transparentcolorbox{Human}{Human} &  \transparentcolorbox{Inline Annotation}{Inline Annotation}& \transparentcolorbox{Indirect}{Indirect} \\ \hline
NotePlayer~\cite{NotePlayer} & UIST & 2024 & \transparentcolorbox{Expand}{Expand} & \transparentcolorbox{Initial}{Initial} & \transparentcolorbox{Human}{Human} &  \transparentcolorbox{Inline Annotation}{Inline Annotation}& \transparentcolorbox{Indirect}{Indirect} \\ \hline
ScatterShot~\cite{Wu2023c} & IUI & 2023 & \transparentcolorbox{Expand}{Expand} & \transparentcolorbox{Initial}{Initial} & \transparentcolorbox{AI}{AI} & \transparentcolorbox{Widget Selection}{Widget Selection} & \transparentcolorbox{Direct}{Direct} \\ \hline
Spellburst~\cite{Angert2023} & UIST & 2023 & \begin{tabular}[c]{@{}l@{}}\transparentcolorbox{Expand}{Expand}\\  \transparentcolorbox{Refine}{Refine}\end{tabular} & \begin{tabular}[c]{@{}l@{}}\transparentcolorbox{Follow-up}{Follow-up}\\  \transparentcolorbox{Initial}{Initial}\end{tabular} & \begin{tabular}[c]{@{}l@{}}\transparentcolorbox{Human}{Human}\\  \transparentcolorbox{AI}{AI}\end{tabular} & \begin{tabular}[c]{@{}l@{}} \transparentcolorbox{Flowchart Manipulation}{Flowchart Manipulation}\\  \transparentcolorbox{Widget Selection}{Widget Selection}\end{tabular} & \transparentcolorbox{Direct}{Direct} \\ \hline
Metaphorian~\cite{Kim2023c} & DIS & 2023 & \transparentcolorbox{Expand}{Expand} & \transparentcolorbox{Follow-up}{Follow-up} & \transparentcolorbox{Human}{Human} &  \transparentcolorbox{Flowchart Manipulation}{Flowchart Manipulation} & \transparentcolorbox{Direct}{Direct} \\ \hline
CoQuest~\cite{Liu2024b} & CHI & 2024 & \transparentcolorbox{Expand}{Expand} & \transparentcolorbox{Follow-up}{Follow-up} & \transparentcolorbox{Human}{Human} & \begin{tabular}[c]{@{}l@{}} \transparentcolorbox{Flowchart Manipulation}{Flowchart Manipulation}\\   \transparentcolorbox{Text Input or Editing}{Text Input or Editing}\end{tabular} & \transparentcolorbox{Indirect}{Indirect} \\ \hline
Sensecape~\cite{Suh2023a} & UIST & 2023 & \transparentcolorbox{Expand}{Expand} & \transparentcolorbox{Follow-up}{Follow-up} & \transparentcolorbox{Human}{Human} & \begin{tabular}[c]{@{}l@{}} \transparentcolorbox{Text Brush Selection}{Text Brush Selection}\\   \transparentcolorbox{Flowchart Manipulation}{Flowchart Manipulation}\end{tabular} & \transparentcolorbox{Indirect}{Indirect} \\ \hline
ExploreLLM~\cite{Ma2024} & CHI EA & 2024 & \transparentcolorbox{Expand}{Expand} & \transparentcolorbox{Follow-up}{Follow-up} & \transparentcolorbox{AI}{AI} & \transparentcolorbox{Widget Selection}{Widget Selection} & \transparentcolorbox{Direct}{Direct} \\ \hline
GAM~\cite{Liu2023e} & CHI & 2023 & \transparentcolorbox{Expand}{Expand} & \transparentcolorbox{Follow-up}{Follow-up} & \transparentcolorbox{AI}{AI} &  \transparentcolorbox{Text Input or Editing}{Text Input or Editing} & \transparentcolorbox{Indirect}{Indirect} \\ \hline
Graphologue~\cite{Jiang2023} & UIST & 2023 & \transparentcolorbox{Expand}{Expand} & \transparentcolorbox{Follow-up}{Follow-up} & \transparentcolorbox{AI}{AI} & \begin{tabular}[c]{@{}l@{}} \transparentcolorbox{Inline Annotation}{Inline Annotation}\\   \transparentcolorbox{Flowchart Manipulation}{Flowchart Manipulation}\end{tabular} & \transparentcolorbox{Indirect}{Indirect} \\ \hline
CoPrompt~\cite{Feng2024} & CHI & 2024 & \transparentcolorbox{Organize}{Organize} & \transparentcolorbox{Initial}{Initial} & \transparentcolorbox{Human}{Human} &  \transparentcolorbox{Tree Manipulation}{Tree Manipulation} & \transparentcolorbox{Direct}{Direct} \\ \hline
VISAR~\cite{Zhang2023b} & UIST & 2023 & \begin{tabular}[c]{@{}l@{}}\transparentcolorbox{Organize}{Organize}\\  \transparentcolorbox{Expand}{Expand}\end{tabular} & \begin{tabular}[c]{@{}l@{}}\transparentcolorbox{Initial}{Initial}\\  \transparentcolorbox{Follow-up}{Follow-up}\end{tabular} & \begin{tabular}[c]{@{}l@{}}\transparentcolorbox{Human}{Human}\\  \transparentcolorbox{AI}{AI}\end{tabular} & \begin{tabular}[c]{@{}l@{}} \transparentcolorbox{Tree Manipulation}{Tree Manipulation}\\  \transparentcolorbox{Widget Selection}{Widget Selection}\end{tabular} & \transparentcolorbox{Direct}{Direct} \\ \hline
CoLadder~\cite{Zhu} & UIST & 2024 & \transparentcolorbox{Organize}{Organize} & \transparentcolorbox{Initial}{Initial} & \transparentcolorbox{Human}{Human} &  \transparentcolorbox{Tree Manipulation}{Tree Manipulation} & \transparentcolorbox{Direct}{Direct} \\ \hline
Low-code LLM~\cite{Cai2024} & NAACL & 2024 & \transparentcolorbox{Organize}{Organize} & \transparentcolorbox{Initial}{Initial} & \transparentcolorbox{Human}{Human} &  \transparentcolorbox{Flowchart Manipulation}{Flowchart Manipulation} & \transparentcolorbox{Direct}{Direct} \\ \hline
Kim et al.~\cite{Kim2023d} & UIST & 2023 & \transparentcolorbox{Organize}{Organize} & \transparentcolorbox{Initial}{Initial} & \transparentcolorbox{Human}{Human} & \begin{tabular}[c]{@{}l@{}} \transparentcolorbox{Flowchart Manipulation}{Flowchart Manipulation}\\  \transparentcolorbox{Widget Selection}{Widget Selection}\end{tabular} & \transparentcolorbox{Direct}{Direct} \\ \hline
PromptChainer~\cite{Wu2022e} & CHI & 2022 & \transparentcolorbox{Organize}{Organize} & \transparentcolorbox{Initial}{Initial} & \transparentcolorbox{Human}{Human} &  \transparentcolorbox{Flowchart Manipulation}{Flowchart Manipulation} & \transparentcolorbox{Direct}{Direct} \\ \hline
DreamSheets~\cite{Almeda2024} & CHI & 2024 & \transparentcolorbox{Refine}{Refine} & \transparentcolorbox{Initial}{Initial} & \transparentcolorbox{Human}{Human} &  \transparentcolorbox{Spreadsheet Manipulation}{Spreadsheet Manipulation} & \transparentcolorbox{Direct}{Direct} \\ \hline
PromptCharm~\cite{Wang2024g} & CHI & 2024 & \begin{tabular}[c]{@{}l@{}}\transparentcolorbox{Refine}{Refine}\\  \transparentcolorbox{Restrict}{Restrict}\end{tabular} & \begin{tabular}[c]{@{}l@{}}\transparentcolorbox{Initial}{Initial}\\  \transparentcolorbox{Follow-up}{Follow-up}\end{tabular} & \begin{tabular}[c]{@{}l@{}}\transparentcolorbox{AI}{AI}\\  \transparentcolorbox{Human}{Human}\end{tabular} & \begin{tabular}[c]{@{}l@{}} \transparentcolorbox{Widget Selection}{Widget Selection}\\  \transparentcolorbox{Visual Brush Selection}{Visual Brush Selection}\end{tabular} & \transparentcolorbox{Direct}{Direct} \\ \hline

\end{tabular}}
  \vspace{-12px}
\end{table*}


\end{document}